\documentclass[journal]{IEEEtran}

\ifCLASSINFOpdf
  \usepackage[colorlinks=true, allcolors=blue]{hyperref}
\else
  \usepackage[dvips]{graphicx}
\fi
\usepackage{url}

\usepackage{graphicx}
\usepackage{amsmath}
\usepackage{booktabs}

\begin{document}

\title{Semantic-Enhanced Feature Matching with Learnable Geometric Verification for Cross-Modal Neuron Registration}

\author{Wenwei Li,
        Lingyi Cai,
        Hui Gong,
        Qingming Luo,
        and Anan Li
\thanks{Under Review. This work was supported in part STI 2030-Major Projects (2021ZD0201002). (\textit{Corresponding author: Anan Li.})}
\thanks{Wenwei Li, Lingyi Cai, Hui Gong and Anan Li are with the Wuhan National Laboratory for Optoelectronics, MoE Key Laboratory for Biomedical Photonics, Huazhong University of Science and Technology, Wuhan 430074, China. (e-mail: liwenwei@hust.edu.cn; lycai@hust.edu.cn; huigong@hust.edu.cn; aali@brainsmatics.org).}
\thanks{Hui Gong and Anan Li are with the HUST-Suzhou Institute for Brainsmatics, JITRI, Suzhou 215123, China.}
\thanks{Qingming Luo and Anan Li are with the Key Laboratory of Biomedical Engineering of Hainan Province, School of Biomedical Engineering, Hainan University, Haikou 570228, China (qluo@hainanu.edu.cn).}
}

\markboth{arXiv Preprint}
{Shell \MakeLowercase{\textit{et al.}}: Bare Demo of IEEEtran.cls for IEEE Journals}
\maketitle

\begin{abstract}
Accurately registering in-vivo two-photon and ex-vivo fluorescence micro-optical sectioning tomography images of individual neurons is critical for structure-function analysis in neuroscience. This task is profoundly challenging due to a significant cross-modality appearance gap, the scarcity of annotated data and severe tissue deformations. We propose a novel deep learning framework to address these issues. Our method introduces a semantic-enhanced hybrid feature descriptor, which fuses the geometric precision of local features with the contextual robustness of a vision foundation model DINOV3 to bridge the modality gap. To handle complex deformations, we replace traditional RANSAC with a learnable Geometric Consistency Confidence Module, a novel classifier trained to identify and reject physically implausible correspondences. A data-efficient two-stage training strategy, involving pre-training on synthetically deformed data and fine-tuning on limited real data, overcomes the data scarcity problem. Our framework provides a robust and accurate solution for high-precision registration in challenging biomedical imaging scenarios, enabling large-scale correlative studies.
\end{abstract}


\IEEEpeerreviewmaketitle

\section{Introduction}
\IEEEPARstart{I}{n} modern neuroscience, one of the central challenges is to accurately track the functional activity of a neuron \textit{in-vivo} and subsequently re-identify its precise anatomical structure in \textit{ex-vivo} tissue samples, thereby elucidating the fundamental structure-function relationship of the brain \cite{markello2022neuromaps, collins2024mapping, wang2022brain}. Two-photon calcium imaging enables high-resolution recording of functional dynamics from neurons deep within living tissue \cite{rosenegger2014high, svoboda2006principles, grienberger2022two}, while fluorescence micro-optical sectioning tomography (fMOST) provides dense anatomical details at nanometer-scale precision, facilitating the reconstruction of neuronal projections across the entire brain \cite{zhong2021high, qi2015fluorescence}. An effective fusion of these two modalities would build a bridge between \textit{in-vivo} function and \textit{ex-vivo} structure, allowing researchers to directly correlate a neuron's functional properties with its fine-grained morphological and even molecular characteristics.

However, achieving precise registration between two-photon and fMOST images presents three formidable challenges. First, acquiring paired datasets is inherently difficult, as it requires imaging the same specifically-labeled neurons in both \textit{in-vivo} and \textit{ex-vivo} states. This laborious, low-yield process results in a critical scarcity of annotated samples. Second, a significant cross-modal appearance difference exists; the images vary immensely in resolution, signal-to-noise ratio, fluorescence intensity, and cellular morphology, rendering traditional registration algorithms based on pixel intensity or low-level textures largely ineffective. Finally, the tissue processing pipeline from the \textit{in-vivo} to the \textit{ex-vivo} state introduces severe and complex geometric deformations that challenge conventional transformation models.

\begin{figure*}[t]
    \centering
    \includegraphics[width=\textwidth]{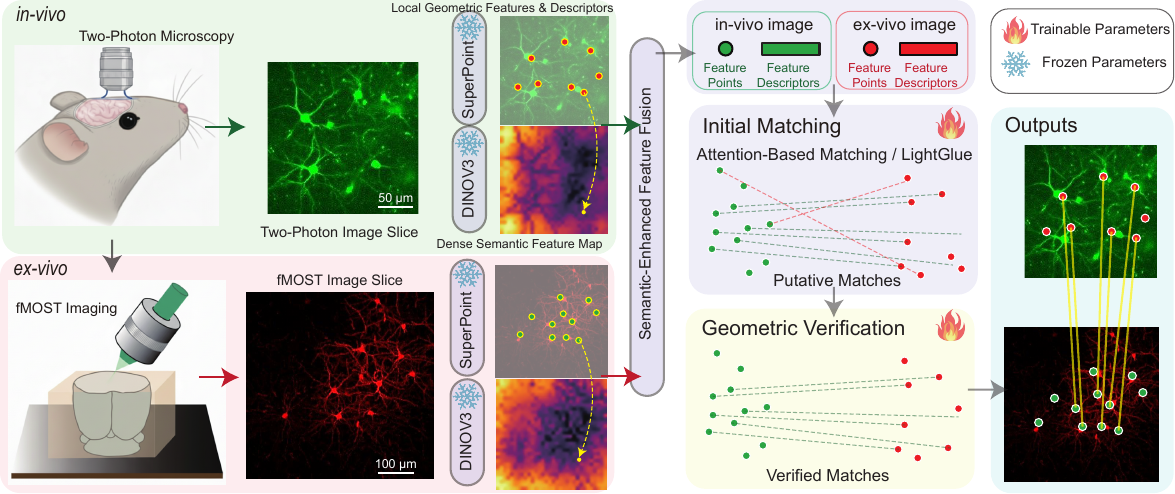}
   \caption{The workflow of our proposed method. Local features from SuperPoint and semantic features from DINOV3 are extracted from \textit{in-vivo} (two-photon) and \textit{ex-vivo} (fMOST) images, respectively, and then fused. Subsequently, LightGlue generates a set of initial matches, which are filtered by a geometric consistency module to produce the final verified correspondences.}
    \label{fig:background}
\end{figure*}

Existing techniques exhibit significant limitations in addressing these challenges. Traditional registration workflows heavily rely on manual expert intervention, such as aligning images by marking specific neurons \cite{wang2022brain} or matching cell clusters through visual inspection \cite{zhou2022mapping}. Such methods are not only highly subjective and time-consuming but also lack the scalability required for large-scale, automated brain mapping initiatives. Some automated approaches, like the graph-based CellGPR \cite{LI2024108392}, model neurons as nodes for matching, but their computational complexity grows exponentially with the number of neurons, making them unsuitable for the dense neuronal scenes captured by fMOST. Furthermore, their sparse node representations fail to accurately describe complex image transformations. On the other hand, mainstream deep learning frameworks like VoxelMorph \cite{balakrishnan2019voxelmorph}, DiffuseMorph \cite{kim2022diffusemorph} and VTN \cite{zhao2019unsupervised}, while capable of learning complex deformation fields, are highly dependent on large-scale training datasets. Thus inapplicable in scenarios characterized by scarce data and complex deformations.

In the field of computer vision, Local feature-based matching methods such as LoFTR \cite{sun2021loftr, bokman2022case} and LightGlue \cite{lindenberger2023lightglue, sarlin2020superglue} have achieved remarkable success in natural image matching tasks within the field of computer vision. However, these models are primarily trained on natural scene data, and the matching strategies they learn do not directly generalize to noisy neuro-optical imaging data. The local feature descriptors upon which they rely face a severe robustness challenge when confronted with the enormous domain gap between two-photon and fMOST imaging.

To address the aforementioned issues, this paper proposes a novel cross-modal image registration framework based on semantic-enhanced local features and a learnable geometric verification module. Our main contributions are threefold:

\begin{enumerate}
    \item \textbf{Semantic-Enhanced Hybrid Feature Descriptor:} We create a novel hybrid descriptor by fusing fine-grained geometric features from SuperPoint \cite{detone2018superpoint} with high-level semantic context from the DINOV3 foundation model \cite{simeoni2025dinov3}, achieving high robustness to cross-modal appearance variations.
    \item \textbf{Learnable Geometric Consistency Module:} We replace traditional RANSAC with a learnable classifier that directly evaluates the physical plausibility of biological tissue deformation. This approach avoids fitting explicit geometric models, making it suitable for non-parametric transformations.
    \item \textbf{Data-Efficient Two-Stage Training Strategy:} To overcome data scarcity, we first pre-train our model on single-modality images with synthetic deformations to learn a general transformation prior, then fine-tune it on a small set of authentic cross-modal data.
\end{enumerate}

\begin{figure*}[t]
    \centering
    \includegraphics[width=\textwidth]{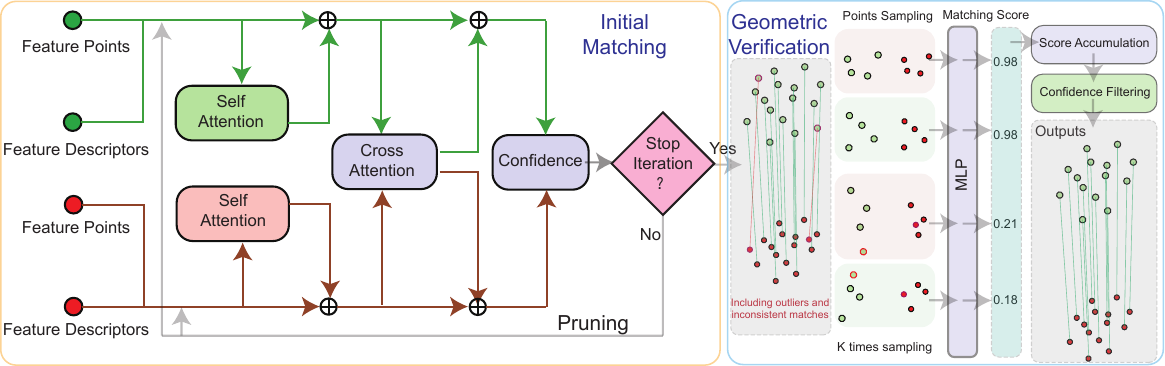}
    \caption{Architecture of the proposed two-stage matching framework. (Left) Initial Matching: A Transformer-based network with self- and cross-attention generates initial correspondences. (Right) Geometric Verification: A learnable module replaces RANSAC by sampling subsets of matches and using an MLP to score their geometric consistency. Outliers are then pruned based on these scores, yielding a final, geometrically coherent set of matches.}
    \label{fig:method}
\end{figure*}

\section{Proposed Method}
\subsection{Overview}

The overall workflow of our proposed method is illustrated in Fig.~\ref{fig:background}. The process begins with extracting local geometric features via SuperPoint and dense semantic features via DINOv3 from both the \textit{in-vivo} source image $I_A$ and the \textit{ex-vivo} target image $I_B$. These multi-level features are then fused to create a robust hybrid descriptor. Subsequently, an attention-based matcher, LightGlue, generates a set of putative correspondences. Finally, a geometric verification module filters these initial matches to yield the final, high-confidence keypoint pairs.

\subsection{Semantic-Enhanced Hybrid Feature Generation}
To overcome the significant appearance gap between two photon and fMOST images, we generate a hybrid feature descriptor for each keypoint that combines geometric localization precision with semantic invariance. Inspired by recent work leveraging vision foundation models to enhance matching generalization \cite{jiang2024omniglue}, we first employ the SuperPoint network to extract keypoint position sets $\mathcal{P}_A = \{p_i^A\}_{i=1}^{N_A}$ and $\mathcal{P}_B = \{p_j^B\}_{j=1}^{N_B}$ from images $I_A$ and $I_B$, respectively, along with their corresponding local geometric descriptors $\mathcal{D}_{\text{local}}^A = \{d_{\text{local},i}^A\}_{i=1}^{N_A}$ and $\mathcal{D}_{\text{local}}^B = \{d_{\text{local},j}^B\}_{j=1}^{N_B}$. Concurrently, we utilize a pre-trained DINOV3 model to extract high-dimensional semantic feature maps, $F_{\text{sem},A}$ and $F_{\text{sem},B}$, from the same images \cite{simeoni2025dinov3}.

For an arbitrary keypoint $p_i^A \in \mathcal{P}_A$, we sample its semantic feature vector $d_{\text{sem},i}^A$ from the feature map $F_{\text{sem},A}$ at the corresponding location using bilinear interpolation, denoted as $d_{\text{sem},i}^A = \mathcal{I}(F_{\text{sem},A}, p_i^A)$. Subsequently, the local and semantic features are concatenated and fused via a small Multi-Layer Perceptron (MLP) to produce the final hybrid feature descriptor $d_{\text{fused},i}^A$:
\begin{equation}
    d_{\text{fused},i}^A = \text{MLP}([d_{\text{local},i}^A ; d_{\text{sem},i}^A])
\end{equation}
where $[;]$ denotes the vector concatenation operation. The same procedure is applied to all keypoints in image $I_B$, yielding the hybrid descriptor sets $\mathcal{D}_{\text{fused}}^A$ and $\mathcal{D}_{\text{fused}}^B$. This fusion mechanism ensures that the final descriptors contain both fine-grained geometric information for precise localization and high-level semantic information to overcome modality differences.

\subsection{Initial Matching and Geometric Verification}
Fig~\ref{fig:method} illustrates the two-stage matching framework. After obtaining robust hybrid feature descriptors, we employ the lightweight feature matcher LightGlue to establish an initial set of correspondences. By processing the descriptor sets $\mathcal{D}_{\text{fused}}^A$ and $\mathcal{D}_{\text{fused}}^B$ through its internal graph neural network with self- and cross-attention mechanisms, LightGlue outputs a set of initial matches $\mathcal{M}_{\text{initial}} = \{(i, j)\}$ with associated confidence scores. However, this initial set is often contaminated with geometrically implausible false matches due to the complex tissue deformations.

To address the fundamental limitation of traditional RANSAC in handling non-parametric transformations, we draw inspiration from the differentiable RANSAC framework \cite{brachmann2017dsac} to design a novel Geometric Consistency Confidence Module (GCCM). The GCCM reframes geometric verification as a classification problem. We randomly sample small subsets of $K=4$ matches, $\mathcal{S}_k$, from $\mathcal{M}_{\text{initial}}$. This sample size provides sufficient robustness for the locally approximate similarity transformation, exceeding the theoretical minimum of two matches. The GCCM module, a function $f_{\text{GCCM}}$, takes only the coordinate information of these match pairs as input and outputs a geometric confidence score $c_k$:
\begin{equation}
    c_k = f_{\text{GCCM}}(\{p_i^A, p_j^B\}_{(i,j) \in \mathcal{S}_k})
\end{equation}
This score quantifies whether the spatial transformation implied by the subset $\mathcal{S}_k$ is biophysically plausible. We then compute an expected confidence score $C(i,j)$ for each initial match $(i,j)$ by averaging the scores of all sampled subsets it participated in. Matches with scores below a predefined threshold $\tau$ are classified as outliers and pruned, yielding the final, geometrically consistent match set $\mathcal{M}_{\text{final}} = \{(i,j) \in \mathcal{M}_{\text{initial}} \mid C(i,j) > \tau\}$.

\begin{figure*}[t]
    \centering
    \includegraphics[width=\textwidth]{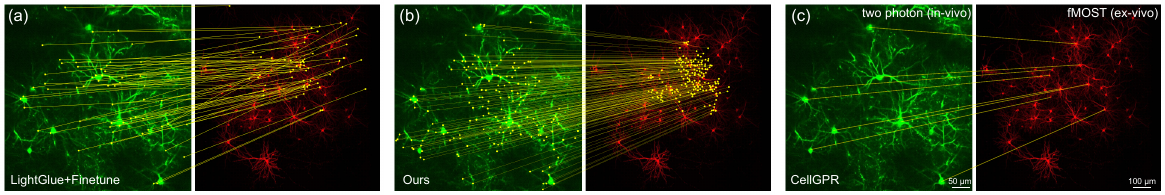}
    \caption{Qualitative comparison of matching results. (a) LightGlue+Finetune produces a moderate number of matches but includes several false positives. (b) Our method generates a dense and highly accurate set of geometrically consistent matches. (c) CellGPR is limited to finding only a few sparse correspondences.}
    \label{fig:qualitative_results}
\end{figure*}

\subsection{Datasets and Training Strategy}
Our research utilizes a dataset of murine brain neurons comprising simultaneously acquired \textit{in-vivo} two-photon calcium imaging and \textit{ex-vivo} fMOST imaging. The dataset originates from 9 sets of fMOST imaging; as a single fMOST set can encompass multiple imaging sites, this yielded a total of 22 two-photon imaging datasets, all of which have been accurately paired with their corresponding fMOST data. The Two-Photon Imaging Data consists of \textit{in-vivo} calcium imaging time-series (x-y-t) from these 22 sites, each containing 1800 frames corresponding to 120 seconds of continuous observation. The fMOST Imaging Data consists of 9 sets of \textit{ex-vivo} tissue blocks covering regions with labeled neurons, at a resolution of $0.65\mu\text{m} \times 0.65\mu\text{m} \times 2\mu\text{m}$, and an approximate volume of $800\mu\text{m} \times 600\mu\text{m} \times 800\mu\text{m}$ each. These 22 accurately matched volumes constitute our ground truth dataset.

\textit{1) Pre-training:} The objective of this stage is to enable the model, particularly the GCCM, to learn a general prior for "plausible deformations." We generate a large-scale synthetic dataset from unpaired single-modality data. Specifically, for an image $I$, we apply a random yet physically plausible non-rigid transformation $T_{\text{synth}}$ (e.g., thin-plate spline) to create a warped version $I' = T_{\text{synth}}(I)$. This provides image pairs $(I, I')$ with dense ground-truth correspondences. We generated over 20,000 such pseudo-matching tasks for initial model training.

\textit{2) Fine-tuning and Testing:} After pre-training, the model is fine-tuned on the real cross-modal data. We partition the 22 ground-truth volumes into 12 for fine-tuning and 10 for testing. To augment the limited fine-tuning set, we generate multiple training instances from each pair by varying projection durations for two-photon data and projection thicknesses for fMOST data. Combined with geometric transformations like rotation, each original pair is expanded into 50 distinct matching tasks, creating a fine-tuning dataset of 600 tasks. During this stage, the parameters of the feature extraction networks (SuperPoint and DINOV3) are frozen, and only the feature fusion MLP, LightGlue, and GCCM are optimized to adapt the model to the specific cross-modal domain gap and real biological tissue deformations.

\section{Experimental Results}

\subsection{Ablation Study}
To validate the effectiveness of the two core components of our framework—the semantic feature fusion and the Geometric Confidence Certification Module —we conducted a series of ablation studies. As shown in Table~\ref{tab:ablation}, the baseline model (SuperPoint+LightGlue with pre-training and fine-tuning) achieved a precision of 46.8\% and a mean Target Registration Error (TRE) of 2.43~$\mu$m.

Upon integrating semantic features, the model's precision increased to 52.8\%, the number of inliers grew by 23\%, and the TRE decreased to 2.26~$\mu$m. This indicates that fusing semantic information effectively bridges the significant modality gap between neuro and optical images, enabling the discovery of more potential correct matches. 
When adding only the GCCM to the baseline model, precision was substantially boosted to 68.4\% and the TRE was reduced to 2.07~$\mu$m. This demonstrates the powerful geometric filtering capability of the GCCM, which efficiently rejects geometrically inconsistent outliers caused by complex deformations.

Finally, our full model, which synergizes the strengths of both modules, achieves the best performance across all metrics: a precision of 74.6\%, a total of 1325 inliers, and a final TRE of 1.68~$\mu$m. These results clearly establish that both proposed modules are crucial for enhancing the accuracy and robustness of cross-modal matching.

\begin{table}[h!]
\centering
\caption{Quantitative Results of the Ablation Study.}
\label{tab:ablation}
\begin{tabular}{@{}lccc@{}}
\toprule
\textbf{Method} & \textbf{Precision} & \textbf{\# of Inliers} & \textbf{TRE ($\mu$m)} \\ \midrule
Baseline & 46.8\% & 684 & 2.43 \\
+ Semantic Features & 52.8\% & 842 & 2.26 \\
+ Geometric Verification & 68.4\% & 1016 & 2.07 \\
\textbf{Ours (Full)} & \textbf{74.6\%} & \textbf{1325} & \textbf{1.68} \\ \bottomrule
\end{tabular}
\end{table}

\subsection{Comparison with State-of-the-Art Methods}
We compared our method against several state-of-the-art feature matching techniques, with results summarized in Table~\ref{tab:sota}. Both LoFTR and the zero-shot LightGlue model performed poorly on this challenging cross-modal task, with precision scores of only 17.8\% and 35.4\% respectively, confirming that general-purpose matchers struggle to generalize to the biomedical domain. Even after fine-tuning, LightGlue's performance, while improved, remained significantly inferior to our method.

The comparison with CellGPR is particularly revealing. As a graph-based method, CellGPR achieves the highest precision (82.6\%) due to its strong global consistency constraints. However, this high precision comes at a steep cost to matching density and computational efficiency. It identified only 58 inliers—a fraction of the 1325 found by our method—which resulted in a slightly higher TRE of 1.79~$\mu$m. Critically, its average computation time was 86.8 seconds, over 200 times slower than our method. This is because CellGPR's computational complexity grows exponentially with the number of nodes, making it impractical for densely featured images.

In contrast, our method achieves the best overall balance. It delivers high precision (74.6\%) while identifying a vastly superior number of inliers (1325), which in turn yields the lowest TRE (1.68~$\mu$m). Furthermore, it accomplishes this with exceptional speed (0.38 seconds). This demonstrates the comprehensive advantages of our approach in terms of accuracy, density, and efficiency.

\begin{table}[h!]
\centering
\scriptsize 
\caption{Comparison with State-of-the-Art Methods.}
\label{tab:sota}
\begin{tabular}{@{}lcccc@{}}
\toprule
\textbf{Method} & \textbf{Precision} & \textbf{\# of Inliers} & \textbf{TRE ($\mu$m)} & \textbf{Time (s)} \\ \midrule
LoFTR (Finetuned)\cite{sun2021loftr} & 17.8\% & 154 & 2.79 & 4.3 \\
LightGlue (Zero-shot)\cite{lindenberger2023lightglue} & 35.4\% & 358 & 2.52 & 0.32 \\
LightGlue (Finetuned)\cite{lindenberger2023lightglue} & 41.5\% & 486 & 2.28 & 0.32 \\
CellGPR\cite{LI2024108392} & 82.6\% & 58 & 1.79 & 86.8 \\
\textbf{Ours (Full)} & \textbf{74.6\%} & \textbf{1325} & \textbf{1.68} & \textbf{0.38} \\ \bottomrule
\end{tabular}
\end{table}

\subsection{Qualitative Results}
Fig. \ref{fig:qualitative_results} provides a qualitative comparison of the matching results applied to two-photon imaging and fMOST imaging. LightGlue (Fig. \ref{fig:qualitative_results}a), while producing a moderate number of matches, includes noticeable false positives, resulting in a cluttered appearance. CellGPR (Fig. \ref{fig:qualitative_results}c) generates only a handful of sparse matches, which is insufficient for accurate transformation estimation. In stark contrast, our method (Fig. \ref{fig:qualitative_results}b) successfully establishes a dense and geometrically coherent set of correspondences between the two images, demonstrating superior robustness and accuracy.

\section{Conclusion}
In this paper, we proposed a novel deep learning framework to address the significant appearance differences and complex deformations in cross-modal neuronal image registration. By fusing local geometric features with global semantic information and employing a learnable geometric verification module, our method generates substantially denser and more precise correspondences, leading to a lower registration error. Our approach strikes a superior balance between accuracy, match density, and computational efficiency, offering a robust and effective solution for large-scale, high-precision correlative analysis between neural function and structure in neuroscience.
\newpage

\bibliographystyle{IEEEtran}
\bibliography{main}
\end{document}